\documentstyle[11pt,newpasp,twoside,epsf]{article}
\def\edcomment#1{\iffalse\marginpar{\raggedright\sl#1\/}\else\relax\fi}
\reversemarginpar

\begin{document}
\title{Geodetic Precession in PSR~B1534+12}
 \author{I. H. Stairs}
\affil{University of Manchester, Jodrell Bank Observatory, Macclesfield, 
Cheshire SK11 9DL UK}
\author{S. E. Thorsett}
\affil{Department of Astronomy and Astrophysics, University of California, 
1156 High St., Santa Cruz, CA 95064 USA}
\author{J. H. Taylor}
\affil{Joseph Henry Laboratories and Physics Department,
       Princeton University, Princeton, NJ 08544 USA}
\author{Z. Arzoumanian}
\affil{NASA Goddard Space Flight Center, Mailstop 662.0, Greenbelt,
	MD 20771 USA}

\begin{abstract}
We present Arecibo observations of PSR~B1534+12 which confirm previous
suggestions that the pulse profile is evolving secularly.  This effect
is similar to that seen in PSR~B1913+16, and is almost certainly due
to general relativistic precession of the pulsar's spin axis.
\end{abstract}

\section{Introduction}\label{sec:intro}

Binary pulsars in close, highly eccentric orbits have long provided
the best strong-field tests of the predictions of gravitational
theories.  Timing observations of PSR B1913$+$16 have allowed the
measurement of three ``post-Keplerian'' parameters: the rate of
periastron advance, $\dot\omega$, the time-dilation and
gravitational-redshift parameter, $\gamma$, and the rate of orbital
period decay, $\dot P_b$ (Taylor \& Weisberg 1989); while PSR
B1534$+$12 permits, in addition, the measurement of the Shapiro-delay
parameters, $r$ and $s$ (Stairs et al. 1998).  These observations have
yielded highly precise tests of both the radiative and quasi-static
predictions of the theory of general relativity.

General relativity also predicts geodetic precession.  The pulsar spin
axis, if misaligned with the orbital angular momentum vector, will
evolve according to:
\begin{equation}
\frac{d{\rm \bf S_1}}{dt} = {\bf \Omega_1^{\rm spin}} \times {\rm \bf S_1},
\end{equation}
where, in general relativity,
\begin{equation}
\Omega_1^{\rm spin} = \frac{G^{3/2}M^{1/2}\mu}{c^2a^{5/2} (1-e^2)}
\left[2+\frac{3}{2}\frac{m_2}{m_1}\right],
\end{equation}
where $G$ is Newton's constant, $M$ is the total mass of the system,
$\mu$ the reduced mass, $c$ the speed of light, $a$ the projected
semi-major axis of the orbit, $e$ the orbital eccentricity, and $m_1$
and $m_2$ the pulsar and companion masses respectively.  For
PSR~B1534+12, $\Omega_1^{\rm spin}$ amounts to $0.52^{\circ}{\rm
yr}^{-1}$; an entire precession cycle would take approximately 690
years.  As the angle between the spin axis and the line-of-sight to
the pulsar changes, the observed cut across the pulsar emission region
will also change, and result in the secular evolution of the pulse
profile.  The long-term change in profile shape of PSR~B1913$+$16 has
been interpreted as evidence for this precession (e.g, Weisberg,
Romani \& Taylor 1989; Kramer 1998; Weisberg \& Taylor, these
proceedings).  Similar profile evolution in PSR~B1534+12 was noted by
Arzoumanian (1995) in examining Arecibo observations from 1990-94;
these changes took the form of an apparent increase with time in the
strength of the interpulse relative to the main pulse at
1400\,MHz.

\begin{figure}
\plotfiddle{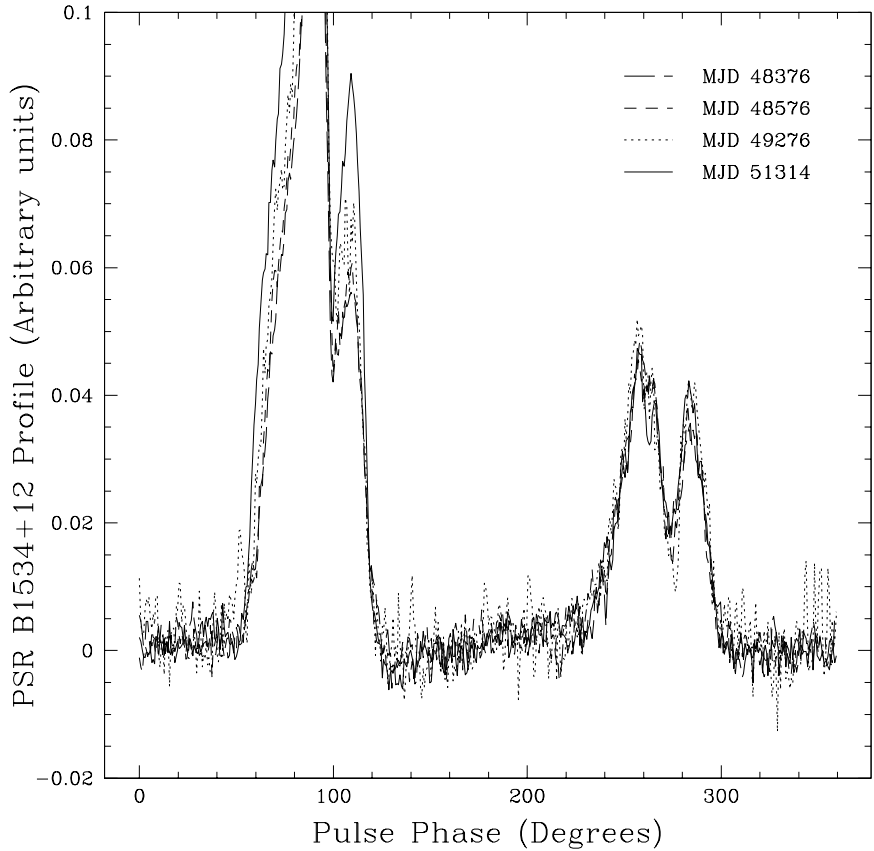}{4.9in}{0}{60}{60}{-180}{-80}
\caption{Evolution of the 1400\,MHz pulse profile of PSR~B1534+12 with time,
using the interpulse as a fiducial reference.} \label{fig:shplot}
\end{figure}

\section{Observations and Results}\label{sec:obs}

\begin{figure}
\plotfiddle{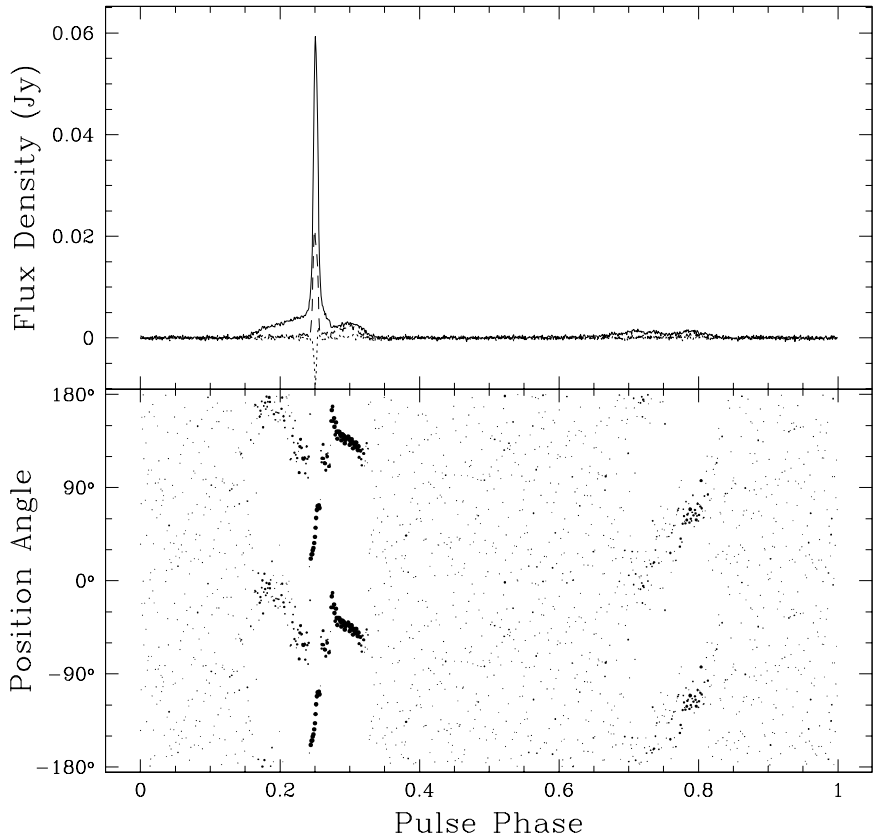}{4.9in}{0}{60}{60}{-180}{-80}
\caption{The 1400\,MHz pulse profile of PSR~B1534+12.  The solid,
dashed and dotted lines indicate total, linear and circular power,
respectively.  The polarization position angle is shown in the lower
panel.}
\label{fig:pol}
\end{figure}

In May 1999, we carried out 1400\,MHz observations of PSR~B1534+12 at
Arecibo Observatory to investigate whether this secular evolution was
still occurring.  During the Arecibo upgrade period, we had developed
a 10-MHz bandwidth coherent-dedispersion baseband recorder, known as
``Mark~IV,'' which we used in parallel with the older $2\times
32\times 1.25$\,MHz ``Mark~III'' filterbank.  This strategy enabled us
to compare our newest profile with those of Arzoumanian, which were
also obtained with Mark~III, to identify any possible small
instrumental differences between the two observing systems, and to
establish a baseline coherently-dedispersed profile against which to
compare future observations.

Our recent data confirm Arzoumanian's observations of precession, and
reveal new detail in the pulse profile evolution.  This can be seen in
Figure~1, in which the small interpulse is taken as a reference point.
The profile dated MJD 51314 is from our May 1999 observations with the
Mark~III system; the earlier three profiles are taken from Arzoumanian
(1995).  It is clear that the ``wings'' of the main pulse are growing
relative to the interpulse, and it appears that the total width at
roughly the half-height of the wings is also increasing with time.
The peak height of the main pulse relative to the interpulse is also
changing, though not monotonically: it decreased between MJDs 48376
and 49276, then began increasing again at some point prior to MJD
51314.

While the change in relative strength of the profile components is
certainly an indication that we are viewing a slightly different part
of the emission region, this cannot by itself provide an estimate of
the change in viewing angle.  However, if the apparent change in width
of the base of the main pulse is real, it will be possible to combine
this information with the magnetic geometry determined from the
Rotating Vector Model (RVM) to put together a model of the beam shape
and change in viewing angle, as has been done for PSR~B1913+16 (e.g.,
Cordes, Wasserman \& Blaskiewicz 1990, Kramer 1998, Karastergiou et
al., these proceedings), and perhaps even to test quantitatively the
rate at which precession is occurring.  The latter may be possible for
PSR~B1534+12 because its magnetic geometry is very well constrained
from polarization observations.  With the Mark~IV instrument, we have
obtained a high-resolution polarization profile of this pulsar at
1400\,MHz (see Figure~2); the resulting fit to the RVM agrees fairly
well with the ``Fit B'' fit derived by Arzoumanian et al (1996), in
which the magnetic inclination angle $\alpha = 114^{\circ}$ and the
impact parameter $\beta = -19^{\circ}$.  If long-term changes in
polarization properties can be measured, or if aberration-induced
orbital effects can be found, we will have a way to determine the
entire orbital geometry of the system, including the vital but
currently poorly-constrained spin-orbit misalignment angle, and
achieve a test of the rate of precession.

\section{Acknowledgments}

We thank Eric Splaver for help with data reduction.

\end{document}